# Water Assisted Proton Transport in Confined Nanochannels

*Xinyou Ma*[1], *Chenghan Li*[1], *Alex B. F. Martinson*[2] *and Gregory A. Voth*[1]\*

[1]Department of Chemistry, Chicago Center for Theoretical Chemistry, The James Franck Institute, and Institute for Biophysical Dynamics, The University of Chicago, Chicago, IL 60637, United States

[2]Materials Science Division, Argonne National Laboratory, Argonne, Illinois 60439, United States

**Abstract**

Hydrated excess protons under hydrophobic confinement are a critical component of charge transport behavior and reactivity in nanoporous materials and biomolecular systems. Herein excess proton confinement effects are computationally investigated for sub-2 nm hydrophobic nanopores by varying the diameters ($d$ = 0.81, 0.95, 1.09, 1.22, 1.36, 1.63, and 1.90 nm), lengths ($l$ ~3 and ~5 nm), curvature, and chirality of cylindrical carbon nanotube (CNT) nanopores. CNTs with a combination of different diameter segments are also explored. The spatial distribution of water molecules under confinement are diameter-dependent; however, proton solvation and transport is consistently found to occur in the water layer adjacent to the pore wall, showing an "amphiphilic" character of the hydrated excess proton hydronium-like structure. The proton transport free energy barrier also decreases significantly as the nanopore diameter increases and proton transport becomes almost barrierless in the $d > 1$ nm nanopores. Among the nanopores studied, the Zundel cation ($H_5O_2^+$) is populated only in the $d$ = 0.95 nm CNT (7,7) nanopore. The presence of the hydrated excess proton and $K^+$ inside the CNT (7,7) nanopore induces a water density increase by 40 and 20%, respectively. The $K^+$ transport through CNT nanopores is also consistently higher in free energy barrier than proton transport. Interestingly, the evolution of excess protonic charge defect distribution reveals a "frozen" single water wire configuration in the $d$ = 0.81 nm CNT (6,6) nanopore (or segment), through which hydrated excess protons can only shuttle via the Grotthuss mechanism. Vehicular diffusion becomes relevant to proton transport in the "flat" free energy regions and in the wider nanopores, where protons do not primarily shuttle in the axial direction.

**Keywords:** proton transport, Grotthuss mechanism, CNT, confinement, free energy

1. **Introduction**

The solvation and mobility of the excess proton in water has been extensively studied due to its importance in numerous areas of chemistry, materials science, and biology.[1-8] The hydration (solvation) structure involves a strongly delocalized excess positive charge density across a core hydronium-like structure and several neighboring water molecules. The solvation environment continuously varies with time, including an interconversion of two limiting hydrated excess proton structures in water: the Zundel ($H_5O_2^+$) and the Eigen ($H_9O_4^+$) cations.[5, 9-13] As the positive excess charge is translocated through the hydrogen bond network, the hydrated excess proton shuttles through neighboring water molecules via successive hopping events and rearranged bonding topologies, which is often referred to as the "Grotthuss mechanism".[14-15] This mechanism is critical to understanding the diffusion and transport of a hydrated excess proton through one-dimensional channels in both biomolecular and material systems. The transport of water and dissolved ions through synthetic and natural membranes is, for example, motivated by the larger effort to improve energy efficiency in water remediation and re-use.

Confinement effects induce atypical chemistry in hydrophobic nanoporous materials.[16-51] Relative to bulk water, important properties including but not limited to spatial and orientation distributions,[23, 26, 30, 32, 37-38, 41, 48-49] thermodynamic properties,[34] nanoscopic water dynamics,[17-18, 26-27, 32, 45, 49] freezing behavior,[24, 39, 44, 47] diffusion rates,[21, 48-49] dielectric constant,[42] and autodissociation,[50] differ significantly in hydrophobic confined pores. Sub-10 nm nanopores (i.e., single-digit nanopores)[51] often show good transport efficiency and ion selectivity.[29, 31] In the smallest hydrophobic nanopore that allow water permeation, water molecules are aligned as a single file

water wire.[16-22, 25-27, 29-30, 32-35, 38, 40, 42-45, 47, 49-51] Even small alterations of the nanopore topology, e.g., the nanopore diameter, often leads to significant changes in the properties of water in these nanopores. Moreover, the nature of the water–hydrophobic interface,[28, 52-53] in addition to confinement effects, also plays an important role in understanding water behavior in hydrophobic nanopores.

The confinement of water in hydrophobic nanopores also dramatically influences proton transport.[16-17, 19-20, 22, 25, 33, 36-37, 40, 43, 54-57] The hydrated excess proton has been predicted to transport through a hydrophobic nanopore via two types of diffusion mechanisms: 1) vehicular diffusion, where the hydrated excess proton structure moves together as a non-reactive hydronium cation by rearranging its solvation structure (standard diffusion), and 2) Grotthuss diffusion, where a reactive proton is shuttled through the hydrogen bond network via hopping to neighboring water.[5, 54-55, 57-58] A carbon nanotube (CNT) with simple cylindrical topology is often used to model a proton channel or ion channel in transmembrane proteins, such as gramicidin A[59] or sarco/endoplasmic plasmic reticulum $Ca^{2+}$-ATPase (SERCA).[60] Almost two decades ago, Brewer et al.[16] first predicted a ~10 times faster than in bulk water proton transport in a narrow nanochannel[16, 61] ($D_{H+}$ = 3.84 ± 0.24 Å$^2$/ps) using the multi-state empirical valance bond (MS-EVB) simulation method.[11, 57-58, 62-63] Dellago et al. later reported a similar proton diffusion constant ($D_{H+}$ = 3.86 Å$^2$/ps) with MS-EVB simulations.[19] Both simulations were equivalent to studying proton transport in a $d$ = 0.8 nm CNT (6,6) nanopore. These were followed by several studies to understand the mechanism of forming a single water wire and proton transport in such a system. For example, Dellago et al.[25] calculated the free energy profile of a hydrated excess proton transport through an array of CNT (6,6) pores, finding a free energy barrier of ~12 kcal/mol (compared to the free energy barrier of a few kcal/mol in bulk water). Cao et al.[33] further revealed a "Zundel–Zundel" proton transport mechanism under confinement through a transient $H_7O_3^+$ structure, which is different from the "Zundel–Eigen–Zundel" mechanism in bulk water.[12] Peng et al.[40] also identified a remarkable "trapping-wetting-permeation" three-step mechanism: 1) a hydrated excess proton diffuses to the water surface and is "trapped" outside the CNT (6,6) entrance in a small free energy minimum, the pore is "dry" until 2) the hydrated excess proton induces water to move into CNT and creates its own water wire, and then 3) the proton transports through the CNT nanopore by shuttling through the intact but transient water wire. More recently, Tunuguntla et al. used 3D fluorescence spectroscopy to measure the temperature-dependent proton transport rate constants using DOPC liposomes embedded CNTs.[43] They reported an experimentally determined Arrhenius activation energy $E_a$ of 13.3 kcal/mole for hydrated excess proton transport through the $d = 0.8$ (CNT (6,6)) nanopore, which is on par with the computational predictions in this work (see later discussion).

In this work we expand the variable space of simulated cylindrical CNT nanopores to include different diameters, lengths, and chirality to investigate the effect of CNT nanopore topology on hydrated excess proton solvation and transport. The MS-EVB 3.2 model[63] is used to perform multi-state reactive molecule dynamics (MS-RMD) simulations[5, 11, 56-58, 62, 64-65] of proton transport through nanopores between graphene sheets. The confinement effects from different CNT nanopores with varying topology are first demonstrated through the internal water and hydrated excess proton distributions. The water and hydrated excess proton lateral radial distributions do not change linearly as the diameter increases, but instead reveal a sharp transition. The evolution of the hydrated excess proton radial distribution, proton solvation structure population, excess protonic charge distribution, and water density provides a comprehensive understanding of proton transport behavior in thus diverse collection of CNTs. The proton permeation free energy profile is also calculated and changes rapidly from the $d = 0.81$ CNT (6,6) to 1.09 nm CNT (8,8). The free energy barrier for CNT (6,6), after accounting for the electrostatic free energy contribution from the lipid bilayer and the activation entropy in the experiment, is thus estimated to be in agreement with the experimental result of Tunuguntla et al.[43]

The charge distribution in the proton solvation structure also reveals the dynamical behavior of the excess proton solvation structure in the CNT nanopores having different diameters. With the evolution of the protonic charge distribution, it is revealed for the first time that 1) in the CNT (6,6), a "frozen" single protonated water wire configuration exists and the Grotthuss diffusion only along the channel axis; 2) in the CNT (7,7), the proton shuttling occurs mostly within the Zundel cation which is, interestingly, perpendicular to the channel axis; and 3) in $d > 1$ nm nanopores, Grotthuss diffusion occurs primarily in the water layer closest to the pore wall. The $d = 0.95$ nm CNT (7,7) nanopore shows a series of interesting features which are significantly different from the other nanopores. Finally, proton transport is compared with $K^+$ transport and we predict the free energy profiles and cationic solvation distributions under confinement to be significantly different.

## 2. Computational Methodology

### 2.1 MS-EVB model

The hydrated excess proton was treated explicitly with the multiscale reactive molecular dynamics (MS-RMD) simulation method developed by the Voth group, using the MS-EVB model (version 3.2)[63] and the SPC/Fw water model.[66] Proton solvation and transport with the MS-EVB model is detailed in previous work;[5, 11-12, 56-58, 62-65, 67] however, a few features are briefly highlighted here. In the MS-EVB model, the reactive potential energy surface (PES) is a linear combination of multiple diabatic state PESs, which represent different configurations of protonation states by defining rearranged hydrogen and chemical bond topologies. A specified protonation configuration, which contains a non-reactive hydronium, is defined by its corresponding diabatic PES and thus the protonic charge is localized in each MS-EVB basis state. Delocalization of the protonic charge in the hydrogen bond network can be interpreted by the weight of the $i^{th}$ MS-EVB state amplitude $c_i^2$, i.e., a fraction of the excess positive charge, where $c_i$ ($i = 1,\ldots, N$) is the ground state MS-EVB coefficient diagonalized from $N$ MS-EVB basis states. The "pivot" hydronium is defined by the MS-EVB state with the largest weight $c_1^2$, which, together with the second-largest weights $c_2^2$, are used to distinguish the limiting solvation structures of a hydrated excess proton. In water, an Eigen cation ($H_9O_4^+$) can be identified by $c_1^2 \sim 0.6$ and $c_2^2 \sim 0.13$ and a Zundel cation ($H_5O_2^+$) by $c_1^2 \sim c_2^2 \sim 0.45$. These values are expected to be similar under confinement, and the difference between the two largest weights $\delta c^2 = c_1^2 - c_2^2 \approx 0.45$ and $\delta c^2 \approx 0$ are used to identify a limiting Eigen and Zundel species, respectively. The effective position of the excess proton charge defect can be defined by the center of excess charge (CEC) position, given by

$$r_{\text{CEC}} = \sum_i c_i^2 \cdot r_{\text{COC},i} \qquad (1)$$

where $r_{\text{COC},i}$ is the hydronium center of charge position of the $i^{th}$ MS-EVB basis state. An extension of the LAMMPS software,[68] called RAPTOR,[67] was used to carry out the MS-RMD simulations. The force field parameters which describe the interaction between a hydrated excess proton and the hydrophobic nanopore, i.e., graphene and the CNT, are consistent with the previous study by Peng et al.[40]

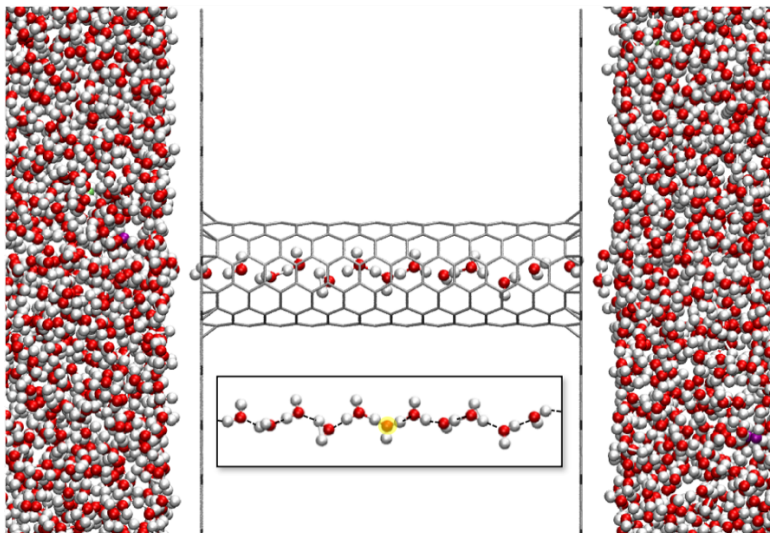

**Figure 1.** A side view of the graphene–CNT (6,6)–graphene nanopore sandwiched by bulk water. A single water wire including a hydrated excess proton traverses the pore and connects bulk waters on the two sides. The orange highlighted excess proton CEC position together with the labeled hydrogen bond network inside the CNT pore is shown in the inset.

In the MS-EVB model, the charge intensity of the $i^{\text{th}}$ MS-EVB state, i.e., the positive charge scaled by $c_i^2$, is a fraction of the protonic charge. The $c_i^2$ and position of the $i^{\text{th}}$ MS-EVB state hydronium, approximated by its oxygen position $\mathbf{r}_{O,i}$, provides an MS-EVB representation of the delocalized protonic charge, i.e., $Q(\mathbf{r}_{O,i}) = c_i^2$. This can be rationalized by the quantum analog that the ground state wavefunction and density of the hydrated excess charge is a linear combination of all diabatic states, respectively. Also, the hydrated excess proton solvation shell structure can be represented by such a protonic charge distribution, where the oxygen positions provide a better description for solvation shell structures by avoiding unnecessary fluctuations introduced by hydrogen atom motions.

## 2.2 CNT nanopores

Different diameters, lengths, curvatures and chirality of CNTs were used as hydrophobic nanopores to investigate the confinement effect of nanopore topologies on hydrated excess proton transport. The nanopore layout, similar to the previous work by Peng et al.,[40] consists of two water slabs separated by a rigid carbon nanopore (see Figure 1). The two water slabs of in total 3028 water molecules collectively contain an excess proton and seven pairs of $K^+$ and $Cl^-$ ions, and the carbon nanopore is constructed with two graphene sheets connected by a carbon nanotube (CNT), where all carbon atoms are kept rigid in all simulations. The armchair style CNT (n,n) (n = 6–10, 12, and 14) with a length of 2.95 nm were used to investigate cylindrical nanopores with different diameters, where $d$ = 0.81, 0.95, 1.09, 1.22, 1.36, 1.63, and 1.90 nm, respectively. In addition, CNT (6,6) with lengths of 2.95 and 5.03 nm were used to study the length effect of cylindrical nanopores. To avoid ambiguity, the center of the CNT in the axial direction and the radial direction are identified hereafter as CNT midpoint and CNT center, respectively.

Three types of additional CNT nanochannel models were used to study nanopores with more complex topologies: 1) a curve-shaped CNT, 2) three "dumbbell"-shaped CNTs, and 3) a "double-gate"-shaped CNT. The curved CNT model has a smooth curvature and the same diameter of CNT (6,6) and a tangential axial length of 3.68 nm, where the distance between the two graphene plates is ~3 nm. A "dumbbell"-shaped CNT consists of three cylindrical CNT segments: two $d$ = 1.09 nm channels and a bottleneck ($d$ = 0.81 nm) in the middle. Three lengths of the bottlenecks are ~0.5, ~1.0, and ~1.5 nm. A "double-gate" nanopore model was used to approximate the inner topology of the transmembrane ion transport channel in a protein such as SERCA. In this latter model, the two $d$ = 0.81 Å bottlenecks symmetrically separate the nanopore into three segments: an entrance channel, an exit channel, and an intermediate channel ($d$ = 1.09 Å). The lengths of the entrance segment are ~5 Å, bottlenecks ~5 Å, and the intermediate channel ~10 Å. In the above nanoporous systems with complex topology CNTs, the distance between two graphene sheets is the same as the systems using cylindrical CNTs with uniform diameters. The various CNTs studied are depicted in various figures below.

2.3   Free energy calculations

Umbrella sampling and weighted histogram analysis method[69] (WHAM) were used to calculate the proton transport free energy profile through the nanotubes, $F(z^+)$, with a bias potential of the form $U_{\text{bias}}(z^+, N_w) = \left(\frac{1}{2}\right) k_z (z^+ - z_0)^2 + U_r(r^+, r_0)$, where $z^+$ is the axial position of the CEC with respect to a CNT pore, i.e., $z^+ < 0$ and $z^+ > 0$ respectively refers to the CEC outside and inside the pore, $r_0$ the CNT radius, and $k_z$ = 10.0 kcal/mol/Å$^2$. The umbrella sampling windows were positioned along the 1D proton transport path in the -5 ≤ $z^+$ ≤ 15 Å region with a 0.5 Å interval, where a 1 ns MS-RMD simulation was performed in each window with a 1.0 fs integration step and the Nosé-Hoover thermostat set to 300 K with a temperature damping parameter of 100 fs. The long-range interactions were treated by the particle–particle, particle–mesh method.[70] Since each CNT is symmetric about its midpoint, the free energy curve beyond the midpoint is expected to be symmetric to that in the $z^+$ ≤ 15 Å region. In addition, the free energy profile of potassium cation transport through CNT was calculated by umbrella sampling simulations with biasing the K$^+$ axial position $z^+$.

To correctly account the free energy difference that reflects the topology of a nanoporous channel, as addressed by Roux and co-workers,[71-72] the radial bias potential

$$U_r = \begin{cases} 0 & r^+ \leq r_0 \\ k_r \cdot (r^+ - r_0)^2 & r^+ > 0 \end{cases} \quad (2)$$

was used to constrain the CEC radial position outside the CNT, where $k_r$ = 10.0 kcal/mol/Å². The CNT radius $r_0$ was used in the cylindrical constraint so that the excess charge radial position is constrained by only the molecular interaction and bias potential inside and outside the CNT, respectively. For the curved shaped CNT, the axial and radial positions were biased along the CNT tangential direction and in the normal plane, and for CNTs with different diameters, the largest radius was used in $U_r$. Before each umbrella sampling simulation, the system was equilibrated in the constant NVT ensemble.

An additional degree of freedom, given by the collective variable (CV) of water occupancy number $N_w$, provides a better understanding of the role of water molecules in proton transport and reactivity. In the previous work by Peng et al.,[40] a two-dimensional potential of mean force (2D-PMF), with the water occupancy as the second CV and the excess proton CEC along the nanotube axis as the first, successfully showed the correlation between the CEC position and the water density and revealed a unique "trapping-wetting-permeation" three-step mechanism. In the free energy calculation of water density inside a CNT, an additional potential function $\frac{1}{2}k_{N_w}(N_w - N_0)^2$ was used to bias the number of water molecules inside CNT with $k_{N_w}$ = 5.0 kcal/mol. $N_w$ the water occupancy number is calculated with the function

$$N_w = \sum_i^{N_{H_2O}} \prod_\alpha^{x,y,z} [1 + R_{i,\alpha}^6]^{-1} \quad (3)$$

And $R_{i,\alpha}$ the dimensionless coordinate is defined as

$$R_{i,\alpha} = \begin{cases} 0 & |r_{i,\alpha} - r_0| - b_\alpha < 0 \\ 1 + (|r_{i,\alpha} - s_{0,\alpha}| - b_\alpha)/\sigma & |r_{i,\alpha} - r_0| - b_\alpha > 0 \end{cases} \quad (4)$$

where $r_{i,\alpha}$ is the $i^{th}$ water molecule oxygen atom position, $s_{0,\alpha}$ the geometric center of a nanopore, $b_x = b_y = d/2$ and $b_z = l/2$ the radius and half-length of a nanopore, respectively, and $\sigma = 5$ Å is a parameter that further defines Eq. 4 with a smooth transition between 0 (water at the water–graphene interface) and 1 (water inside the nanopore).

## 3. Results and Discussions

### 3.1 Effect of CNT nanopore diameter

#### 3.1.1 Lateral radial distribution of the CEC and water in CNT.

The most intuitive way of defining the effect of confinement in nanopores is through the configurational radial distribution. As depicted in Figure 2a-d and Figure 3a-c, the increase in CNT diameter leads to different water radial distributions, which are in good agreement with those reported previously with[22, 26, 32, 48-49] and without a hydrated excess proton.[23, 26, 30, 48-49] As the hydrated excess proton moves inside the CNT, the solvation structure is strongly confined by the CNT topology. The radial density distribution $\rho(r)$ of the CEC in the lateral direction, i.e., the normal plane perpendicular to the axial direction, can be calculated by $\rho_{CEC}(r) = c \cdot P_r/(2\pi|r|)$, where $P_r$ is the radial population of the CEC at distance $r$ from CNT axis, and $c = 1/\int P_r/(2\pi|r|)dr$ the normalization factor. Similarly, the radial density distribution could be obtained respectively for all hydrogen and oxygen atoms inside the CNT. The radial distributions of the excess proton CEC, hydrogen, and oxygen atoms are calculated as the CEC axial position is located at the CNT midpoint. In the $d = 0.81$ nm CNT (6,6) nanopore, hydrogen and oxygen atoms have a trimodal and bimodal distribution, respectively. This indicates that each water molecule in the single water wire structure donates and accepts one hydrogen bond simultaneously, leaving the outward hydrogen atom "dangling" from the hydrogen bond network, which agrees with the experimental result of Bernadina et al.[41] In the $d = 0.95 – 1.22$ nm nanopores, i.e., CNT (n,n) (n = 7-9), the water and excess proton radial distributions become different (Figure 2 f-h), where water has almost no population at the center and is mostly distributed near the CNT wall (the water "wets" the hydrophobic CNT wall). A small fraction of hydrogen atoms are closer to the CNT wall, which corresponds to outward O–H bonds. Similar to water molecules inside the CNT, the radial distribution of the hydrated excess proton bifurcates and overlaps with the inward hydrogen atoms that participate in the hydrogen bond network. The excess proton CEC distributions, compared to the inward hydrogen atom distributions, are slightly closer to the pore center, which shows that the core or "pivot" hydronium structure donates a hydrogen bond that bridges multiple water chains. Compared to the uniform distribution of the proton CEC in bulk, as shown in Figure 2i at $z^+ = -5$ Å, the above CEC distributions in the CNT show a clear confinement effect.

An interesting question is the correlation between the radial distributions of the water molecules and the hydrated excess proton. In larger CNT nanopores, does the increase in water population at the CNT center lead to a larger excess proton population there? To answer this question, proton solvation and transport was studied with CNT (n,n), n = 10, 12, and 14 nanopores, which have increasing diameters of 1.36, 1.63, and 1.90 nm, respectively. As shown in the water radial density distributions (see Figure 3d-f), a second water layer emerges at the CNT center, which leads to not only intra-layer but also inter-layer hydrogen bond networks. However, in these CNT nanopores, the increase in excess proton probability in the central region is much less significant, and the hydrated excess proton retains a dominant distribution in the layer of water molecules closest to the CNT wall.[53] The analysis of these distributions is two-fold. First, the pivot hydronium is predominantly involved in the intra-layer rather than the inter-layer hydrogen bond network. Second, the confinement effect in a CNT nanopore leads to a significantly higher proton concentration at the water layer closest to the CNT wall, which creates a pH gradient towards the nanopore center. These are consistent with the higher distribution of hydrated excess proton at the water–hydrophobic interface[53] and "amphiphilic" character of the hydrated excess proton.[52-53, 73-74]

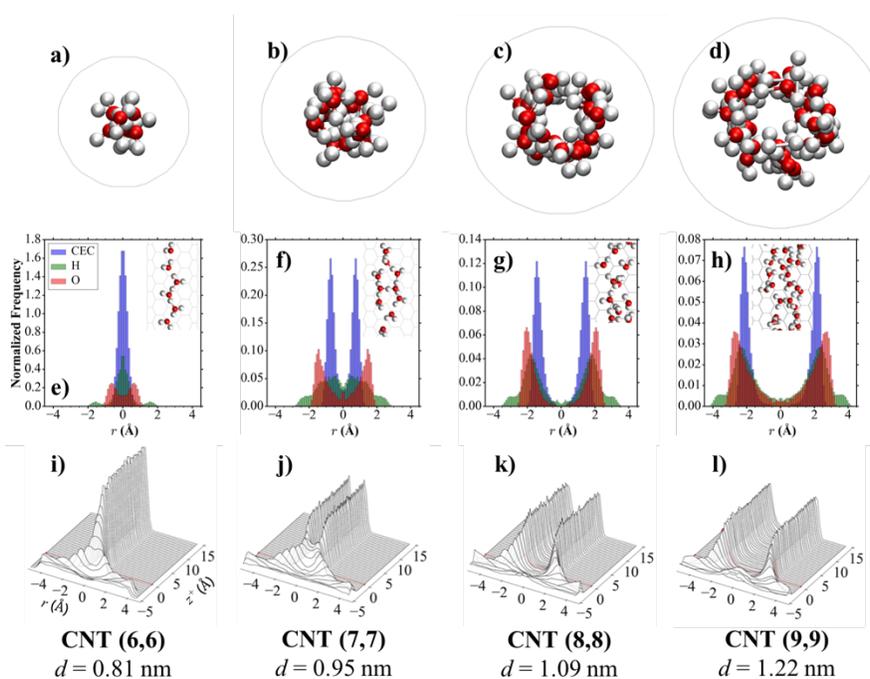

**Figure 2.** a) – d) Top view of water and the hydrated excess proton in CNTs; e) – h) the normalized radial density distribution of the excess proton CEC (blue), hydrogen atom (green), and oxygen atom (red) positions at $z^+ \sim 15$ Å, and i) – l) series of the CEC normalized radial density distributions

at $5 < z^+ < 15$ Å. These are plotted for proton solvation and transport in the CNT (6,6), CNT (7,7), CNT (8,8), and CNT (9,9) nanopores, respectively.

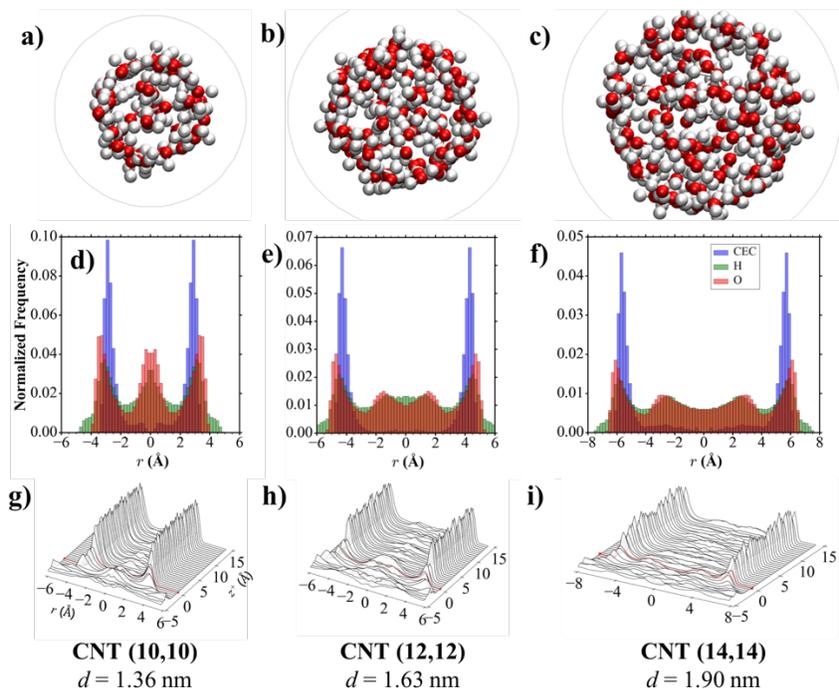

**Figure 3.** Same as Figure 2 but for proton solvation and transport in CNT (10,10), CNT (12,12), and CNT (14,14), respectively.

The water and hydrogen bond networks are determined by the confinement effect and not greatly perturbed by the presence of a hydrated excess proton. In the water layer closest to the CNT wall, water O-H bonds which do not participate in the hydrogen bond network often point outwards rather than inwards, which is consistent with the water orientations by *ab initio*[30, 37] and classical MD simulations.[23, 26, 32, 38, 48-49] Thus, the CNT nanopore "effective radius" $r_{eff}$ can be defined by the outmost hydrogen radial positions according to Figure 2e-h and Figure 3d-f, and this is approximately 0.2, 2.8, 3.0, 4.2, 5.0, 6.2, and 8.0 Å for CNT (n,n) (n = 6, 7, 8, 9, 10, 12, and 14), respectively. Such an effective radius helps to compare early work using effective confining potentials to recent simulations which often use the CNT (n,m) notation and explicit water–CNT interactions. In the same water layer, the hydronium constantly donates three hydrogen bonds, maintaining an orientation with oxygen and hydrogen outwards and inwards, respectively. This is revealed by both the trajectory snapshots and the excess proton CEC and O radial positions. The CEC and O distributions are similar to those of a hydrated excess proton at a water–hydrophobic interface,[53] which is retained regardless of the sharp phase transition in increasing diameters of CNT, indicating that the hydronium orientation is governed by the presence of the hydrophobic interface.

While the radial distribution inside CNT described above is helpful to understand the local structure of the hydrated excess proton under strict confinement, a series of such distributions as the proton transitions from bulk into the CNT (see Figure 2i-l and Figure 3g-i) reveals the subtleties of the excess proton CEC position under partial confinement. First, before proton transport to the CNT nanopore, e.g., the CEC axial position at $z^+ \sim -5$ Å, the hydrated excess proton is evenly distributed in water bulk in the absence of confinement. Second, as a proton approaches the CNT nanopore, the proton population increases near the edge of a nanopore, which shows that the confinement effect is already important before the excess proton is inside a nanopore. In $d > 1$ nm CNT nanopores, the proton population increases in a minor fashion in the central region due to the change in water distribution. Third, as a proton moves inside a CNT nanopore, i.e., $z^+ > 0$, the excess proton radial distribution becomes constant, which implies a different state of the proton solvation structure than in bulk water.

3.1.2  Water orientation and hydrogen bond network distribution.

The orientation of a water molecule can be determined by $\phi$ the angle between the water dipole moment vector $\boldsymbol{\mu}$ and the vector $\mathbf{r}(O_W - O_{TH})$ which is a vector between the oxygen atom positions in this water and the pivot hydronium, i.e.,

$$cos\phi = \frac{\boldsymbol{\mu} \cdot \mathbf{r}(O_W - O_{TH})}{|\boldsymbol{\mu}||\mathbf{r}(O_W - O_{TH})|} \quad (5)$$

As $cos\phi$ approaches a value of 1, $\boldsymbol{\mu}$ aligns with the vector $\mathbf{r}(O_W–O_{TH})$, and $cos\phi$ towards 0 or -1 represents that $\boldsymbol{\mu}$ minimally or "anti-" aligned with $\mathbf{r}(O_W–O_{TH})$, respectively. The distributions of $cos\phi$ shown in Figure 4a reflects water orientation in CNT nanopores with different diameters. In sub-1 nm nanopores, water molecule orientations are overall aligned with the CNT axis towards the hydronium, which is consistent with the strongly confined water orientations in single and double water wires in CNT (6,6) and CNT (7,7). This result agrees with the dipole wire dipole model by Köfinger et al.[32] As the diameter of nanopores increases, $cos\phi$ becomes randomized and the confinement on water orientations decreases.

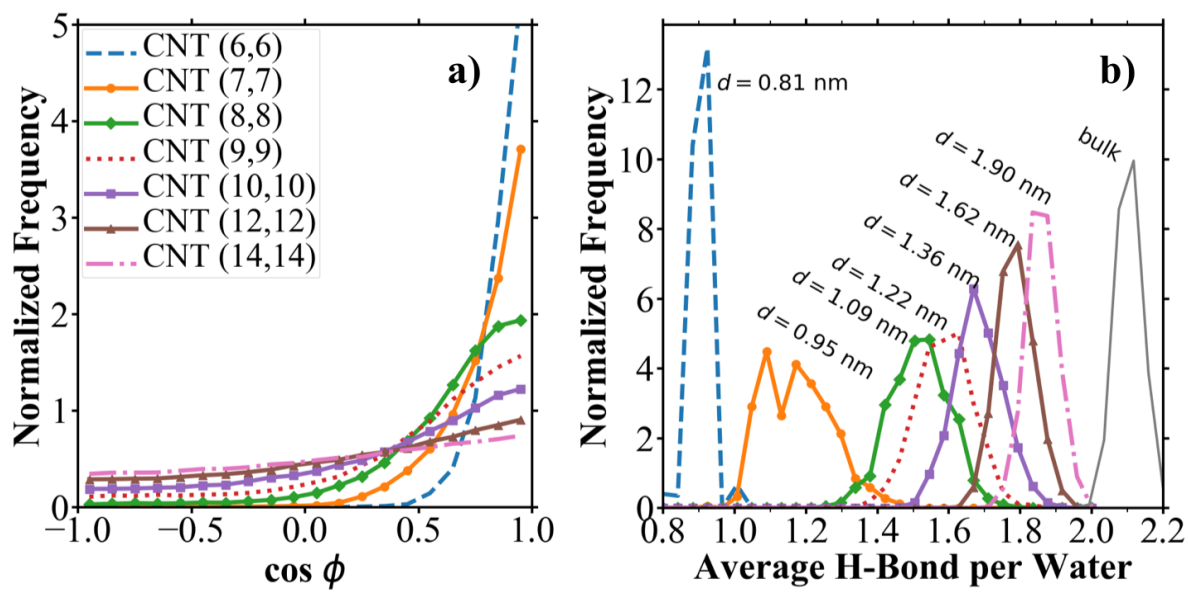

**Figure 4.** The water molecule distributions of a) cos $\phi$, a parameter representing the water alignment (eq. 5), and b) the distribution of the average ratio of the number of hydrogen bonds, formed between water molecules inside each nanopore and in bulk, to number of water molecules. The two distributions are plotted for CNT (6,6) (blue dashed line), CNT (7,7) (orange solid line and dots), CNT (8,8) (green solid line and diamonds), CNT (9,9) (red dotted line), CNT (10,10) (purple solid line and squares), CNT (12,12) (brown solid line and triangles), CNT (14,14) (pink dash-dotted line), and bulk (gray solid line).

The water molecule orientations may also be reflected by the hydrogen bond network inside the nanopores, and the distribution of the average number of hydrogen bonds formed between all water molecules in a nanopore, Figure 4b. In CNT (6,6), each water molecule forms one hydrogen bond, i.e., donating and accepting a hydrogen bond simultaneously, and the narrow distribution agrees with the water alignment in a single water wire. The hydrogen bond network in wider nanopores becomes more complex and the average hydrogen bond per water molecule increases. Particularly, in the $d = 0.95$ nm CNT (7,7) nanopore, this becomes a broad and bimodal distribution, which is explained in detail in Section 3.1.5. Compared to water molecules in the bulk, which forms two hydrogen bonds per water molecule on average, the confinement effect in larger diameter nanopores decreases but does not completely vanish.

### 3.1.3   Zundel and Eigen cation populations

The solvation structure of a hydrated excess proton is correlated to the distribution of solvent molecules. Due to the hydrophobic confinement, the hydrated excess proton may be solvated partially, or even poorly, in a nanopore. In the $d = 0.81$ nm CNT (6,6) nanopore, the two limiting solvation structures are the $H_7O_3^+$ cation and a "distorted" Zundel cation (see Figure 5a). An $H_7O_3^+$ cation and the distorted Zundel cation resembles aspects of an Eigen cation and a Zundel cation with $\delta c^2 \sim 0.45$ and $\delta c^2 \sim 0.0$, respectively, whose populations can be represented by the distribution of $\delta c^2$ (see discussion preceding eq. 1). Each $\delta c^2$ distribution was calculated from an umbrella sampling window with respect to an average excess proton CEC position $z^+$. At different $z^+$, a series of $\delta c^2$ distributions were plotted to show the evolution of such distributions during the proton transport (see Figure 5a). First, as a hydrated excess proton moves from bulk water, the primary solvation structure is the Eigen cation. Second, as it further moves into the CNT mouth, the Eigen cation becomes more populated, among which two water molecules connect to bulk water and the third water connects the single water wire. Third and finally, $H_7O_3^+$ becomes the predominant solvation structure as the hydrated excess proton enters the CNT.

The evolution of the proton solvation structure population is dramatically different for $d = 0.95$ nm, i.e., CNT (7,7). As shown in Figure 5b, the $\delta c^2$ distributions has less fluctuation as the proton moves into the CNT, and, most importantly, the Zundel cation becomes the predominant solvation structure as a proton moves inside the CNT. Interestingly enough, as depicted in Figure 5b by a trajectory snapshot, the Zundel cation is in the *perpendicular* orientation to the CNT axial direction and is stabilized by the double water wire formed on each side.

In the $d > 1$ nm CNT nanopores, the Eigen cation has a higher population throughout the proton transport process, which is due to the increased number of water molecules in the confined space. Since the excess proton CEC has a preferential radial distribution at the water layer closest to the CNT wall, as shown in Figure 2e-h and Figure 3d-f, the Eigen cation formed in the $d > 1$ nm CNT nanopores is mostly distributed at the water–CNT interface. By comparing the diameter-dependent $\delta c^2$ distributions, the $d = 0.95$ nm CNT (7,7) shows a unique feature of promoting the Zundel cation population inside the nanopores. An experimental observation of this remarkable predicted structures would be a valuable target for future research.

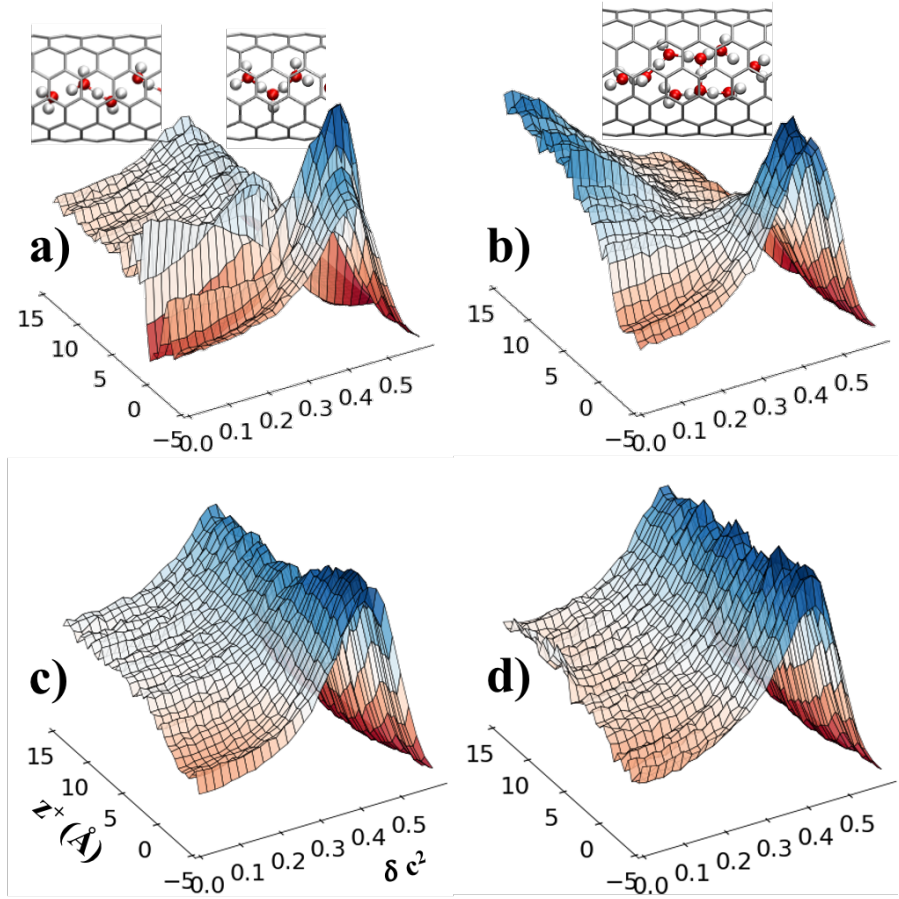

**Figure 5.** Four series of $\delta c^2$ distributions at $5 < z^+ < 15$ Å, indicating the population of the Zundel cation (or the distorted Zundel cation) at $\delta c^2 \sim 0$ and the Eigen cation (or the $H_7O_3^+$ cation) at $\delta c^2 \sim 0.45$ in the a) CNT (6,6), b) CNT (7,7), c) CNT (8,8), and d) CNT (9,9) nanopores, respectively. Trajectory snapshots of a distorted Zundel cation and an $H_7O_3^+$ cation are shown as insets in a) and a Zundel formed in CNT (7,7) as the inset of b).

3.1.4 Pivot hydronium shape

It is also of interest to examine the effect of confinement on the shape of pivot hydronium cation, which can be quantified by an average of the three H-O-H angles, i.e.,

$$\bar{\theta} = \frac{1}{3}\sum_{i=1}^{3} \theta_{HOH,i} \qquad (6)$$

The comparison of $\bar{\theta}$ distribution in CNT nanopores with different diameters is shown in Figure 6. In most CNTs, $\bar{\theta}$ is in the range of 95° to 117° and the most probable H-O-H angle is at 106°. For $\bar{\theta}$ in CNT (6,6) the most probable H-O-H angle is at 112° and has the distribution shifted towards 120°, which corresponds to a planar hydronium. This can be compared to a hydronium cation in the gas phase[75] and aqueous phase,[76] where the equilibrium H-O-H angle is approximately 113.6° and 106.7°, respectively. This result suggests that the protonic charge is more localized in the pivot hydronium of an $H_7O_3^+$ cation, which in turn is less solvated compared to the solvation structures in the CNT nanopores with larger diameters.

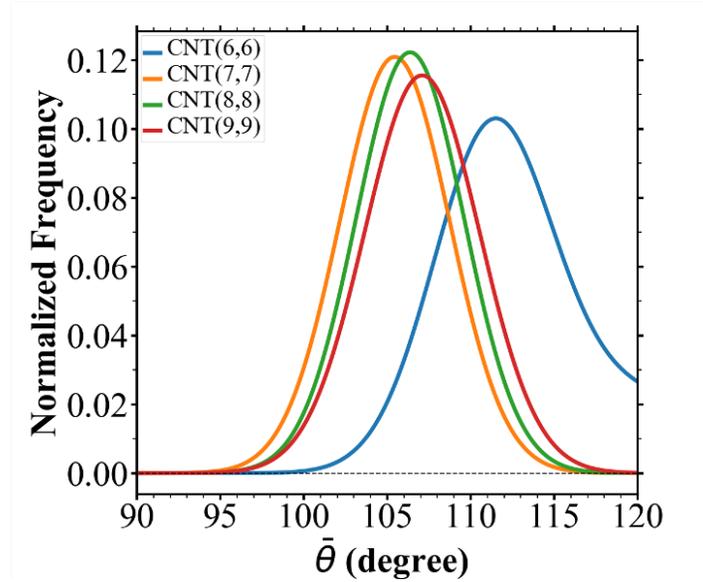

**Figure 6.** The distribution of hydronium shape fitted by gaussian functions.

### 3.1.5 Excess proton induced water density increase

In most of the nanopores, the water distribution under confinement is not correlated to the excess CEC axial position; however, in the CNT (7,7) nanopore, the water molecule radial distribution inside the CNT evolves with the proton axial position. Before an excess proton approaches to the mouth of the nanopore, the trimodal hydrogen atom distribution is similar to that of the single water wire in CNT (6,6) (see Figure 7a), where the hydrogen atoms have a larger population at the center. As the excess proton moves into the CNT, Figure 7b, the central hydrogen distribution bifurcates, and the oxygen atom has a smaller population at the center, which resemble the water distributions in CNT (8,8) and CNT (9,9).

Along with proton transport into the nanopore, a significant increase in water density in the confinement space is found in the $d = 0.95$ nm CNT (7,7) nanopore. Before the proton approaches the nanopore, a single file water wire is established inside CNT (7,7) as shown by a trajectory snapshot in Figure 7c. Such a single water wire is less ordered in CNT (7,7) than in CNT (6,6) and statistically consists of more molecules: the most probable $N_w$ is 13 in CNT (6,6) and 15 in CNT (7,7). As a proton moves into the CNT (7,7) pore, a second water wire is abruptly established next to the first water wire (see Figure 7d). The double water wire increases the complexity of the hydrogen bond network and stabilizes the excess protonic charge. The perpendicularly-oriented Zundel cation, which has a uniquely higher population in CNT (7,7) than other CNT nanopores, is favored by both the particular nanopore diameter and the four surrounding water molecules in double water wire configuration.

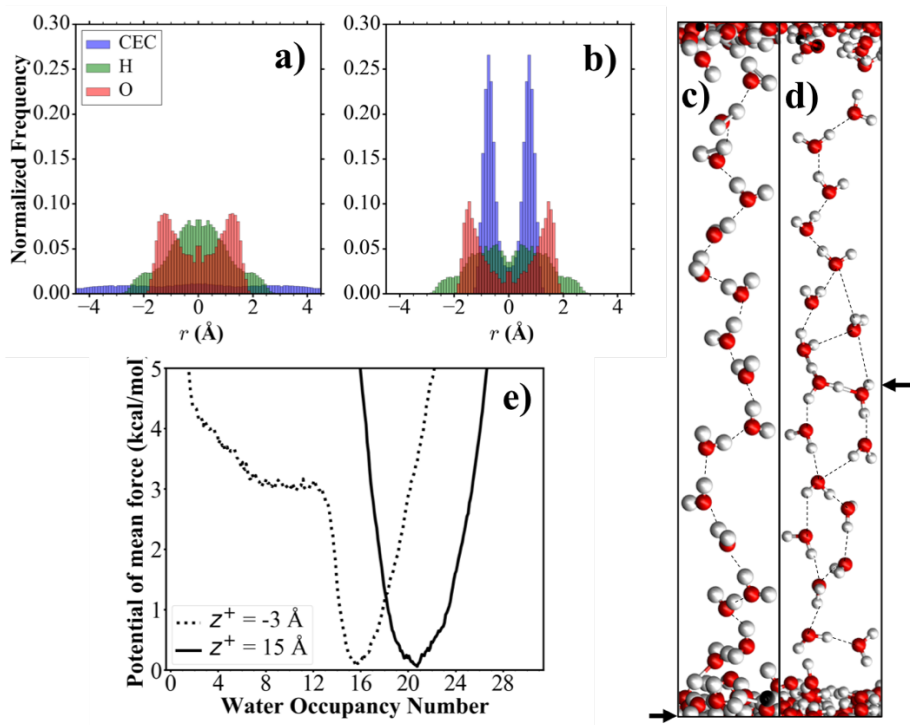

**Figure 7.** a) – b) The radial density distributions of the excess proton CEC (blue), hydrogen atom (green), and oxygen atom (red) positions, c) – d) trajectory snapshots of water molecules inside CNT and hydrated excess proton, where the CEC z positions at -3 and 15 Å are labeled with arrows, respectively, and e) the free energy curves with respect to number of water molecules inside the $d$ = 0.95 nm CNT (7,7) nanopore at $z^+$ ~-3 and ~15 Å, respectively.

This proton-induced water density increase was further quantified through free energy calculations. Umbrella sampling simulations were performed with biasing the number of water molecules inside the pore. At $z^+ \sim -3$ Å, i.e., the excess proton is 3 Å outside the pore, the free energy profile has a minimum at $N_w \sim 15$, and at $z^+ \sim 15$ Å, i.e., the proton is at the center of CNT pore, the free energy has a minimum at $N_w \sim 21$. These free energy profiles are shown in Figure 7e, which indicates that proton inside the nanopore increases the number of water molecules inside the nanopore by almost 40%. Such a significant increase in water density is found exclusively in the CNT (7,7) pore and provides an explanation to the broad and bimodal distribution in Figure 4b. In the other CNT nanopores, e.g., CNT (6,6) and CNT (8,8), the presence of an excess proton in the nanopore only increases the number of water molecules by 1 and 2, respectively. In even wider nanopores, such effect is even less significant.

3.1.6 Free energy of hydrated excess proton transport through CNTs

The free energy profile of hydrated excess proton transport through sub-2 nm hydrophobic nanopores $F(z^+)$, as shown in Figure 8, quantifies the confinement effect in the axial directions of nanopores with increasing diameters. For each nanopore, $F(z^+)$ is plotted for a proton transport from bulk to the nanopore midpoint. In a sub-1 nm nanopore, $F(z^+)$ reaches a minimum at 3.6 Å just outside the CNT mouth, i.e., the water surface at the water–graphene interface, followed by a rapid increase at $-1.0 < z^+ < 3.0$ Å. The free energy curve reaches a plateau as the proton moves to the CNT midpoint, which has a value of 20.8 kcal/mol in the $d = 0.81$ nm CNT (6,6) pore and 8.8 kcal/mol in the $d = 0.95$ nm CNT (7,7) pore. In the CNT (8,8) and CNT (9,9) nanopores, $F(z^+)$ at the CNT midpoint is respectively 1.0 and 1.7 kcal/mol lower than the free energy minimum at $z^+ = -3.6$ Å, which leads to a "barrierless" proton transport process. Such behavior can be explained by the amphiphilic character of the hydrated excess proton that is preferentially distributed at the water–hydrophobic interface as well as the more structured hydrogen bonded network in the CNT relative to the bulk. For CNT nanopores with larger diameters, the free energy profile is very similar to that of CNT (9,9). As the nanopore diameter increases from 0.81 to 0.95 nm, the free energy barrier rapidly decreases from 20.8 to 8.8 kcal/mol. Such a 12 kcal/mol decrease in the free energy barrier is mainly due to 1) the increasing number of water molecules in the CNT and 2) less confinement on the hydrated excess proton via the change in CNT inner curvature. To gain a better understanding of the two mechanisms, a free energy calculation was performed with only water–CNT interaction, i.e., the interaction between CNT and the pivot hydronium were artificially turned off, and the free energy barrier then becomes 10.3 kcal/mol (see Figure 8). Thus, the increase in nanopore diameter from 0.81 nm to 0.95 nm nanopore stabilizes the hydrated excess proton by 12 kcal/mol, nearly 88% of which is contributed by the increase in solvent molecules and 12% by less confinement of the hydronium directly through the hydronium-CNT interaction. In addition, the hydronium shape distribution in Figure 6 shows that the free energy barrier is not correlated with the distortion of $H_3O^+$.

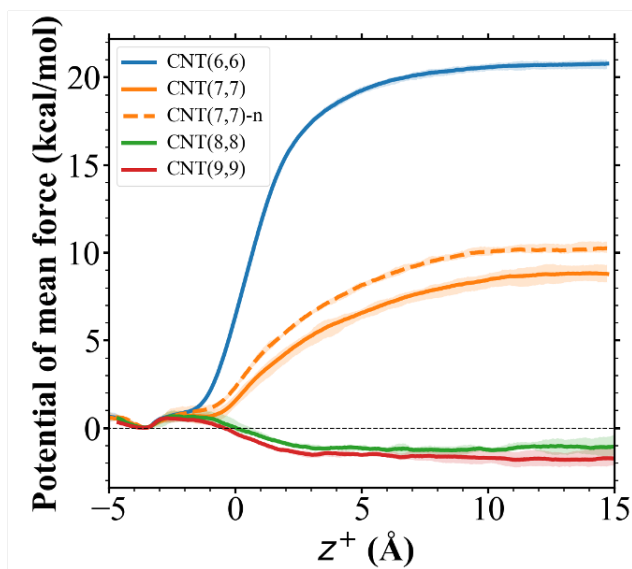

**Figure 8.** the free energy profile of hydrated excess proton transport through sub-1.5 nm nanopores. The free energy curve in the dashed line corresponds to proton transport in CNT (7,7) without hydronium–CNT interactions.

The experimentally measured Arrhenius activation energy $E_a$ of proton transport through $d = 0.8$ CNT nanopores has been reported by Tunuguntla et al., and this values is 13.3 kcal/mol.[43] In their experiment, the temperature-dependent fluorescence decay of pyronine dye encapsulated in a liposome was measured, where several CNTs were inserted into the lipid bilayer membranes as proton transport nanopores. Compared to the experimental activation energy, our MS-RMD simulation reported here focuses on the change in free energy along the 1-D channel, but an experimentally measured activation energy will not include the activation entropy effect of a hydrated excess proton moving from water in the region of the channel mouth into the nanopore (this effect is in the prefactor of the rate expression). In the sub-1 nm CNT nanopores, as the free energy barrier governed by the confinement effect becomes important, the comparison between the experimentally measured Arrhenius activation energy and free energy barrier determined by simulation thus becomes more difficult. Approximately, the experimental proton transport free energy barrier in the experimental $d = 0.8$ nm nanopore can be estimated as $F' = E_a - T\Delta S^{\#} \cong F_{calc}(z^+ = \text{midpoint}) - F_{electro}$, where $\Delta S^{\#}$ is the proton transport activation entropy term and $F_{electro}$ is an electrostatic correction between the excess proton charge and the lipid membrane environment (which has an unknown effective dielectric constant). The difference between the measured 13.3 kcal/mol Arrhenius activation energy and the calculated 20.8 kcal/mol free energy barrier is threefold: First, unlike the nanopores in the current work which are surrounded by vacuum, the lipid bilayers membrane used in the experiment provides stabilization to the hydrated excess charge and thus lead to a lower free energy barrier. With a lipid dielectric constant of $\varepsilon = 2$, the electrostatic energy cost of a cation transporting through a $d = 0.6$ nm nanopore in a lipid membrane was calculated by Levitt and is $F_{electro} = 4.0$ and 6.0 kcal/mol for $l = 2.5$ and 5.0 nm, respectively.[77] This can be used to estimate the $F_{electro}$ in the current study. Second, the proton permeation into the CNT is an unfavorable entropic process, since the CEC is uniformly distributed outside the nanopore but highly localized at the water–CNT interface inside the nanopore. Such an entropy-decrease process results in a positive $-T\Delta S^{\#}$ term of at least several kcal/mol. As a result, the experimental value of $F'$ would be expected to be greater than 13.3 kcal/mol. In addition, the influence of the carboxyl groups at the end of CNTs on the Arrhenius activation energy is unknown. Thus, the experimental free energy barrier of proton transport through a $d = 0.8$ nm CNT (6,6) is estimated to be approximately in the range of 20 kcal/mol, in agreement with our simulated value.

3.1.7 Excess protonic charge distributions

The distribution of an excess protonic charge across hydrating water molecules, as described by the MS-EVB model, can be represented by the intensity of the protonic charge in each solvent molecule. For water molecules in the excess proton solvation structure, each has a partial protonic charge $Q(\mathbf{r}_{O,w}) = \delta(\mathbf{r}_{O,i} - \mathbf{r}_{O,w}) \cdot c_i^2$, respectively, where $\mathbf{r}_{O,w}$ is the position of oxygen atom. For water molecules in the constant NVT ensemble, the protonic charge distribution, calculated from the sampled solvation structures at a specific excess proton CEC axial position, is more meaningful for proton transport through nanopores. In the absence of such confinement, the hydrated excess proton maintains a full hydration sphere and the proton charge is strongly delocalized. Subject to the water distribution in bulk, the protonic charge distribution is also spherically dispersed, and the intensity decreases at increasing distance from the CEC. However, the water molecule distribution under confinement, especially in a sub-1 nm pore, is subject to the nanopore topology and hence significantly changes the protonic charge distribution. As an excess proton approaches the nanopore from the bulk, a portion of its solvent shell is in the nanopore. The protonic charge distribution under confinement is thus distinct from that in the bulk, and the contrast is shown in Figure 9I-A and Figure 9II-A.

A series of protonic charge distribution snapshots in Figure 9 illustrates proton transport in the axial direction. In a sub-1 nm nanopore, the protonic charge becomes less delocalized due to a reduced number of solvent molecules and is only delocalized along the single water wire (see Figure 9I). In the $d = 0.81$ nm CNT (6,6) nanopore, at $z^+ \sim 0$, the protonic charge has a high intensity on the core (or pivot) hydronium, which corresponds to the higher $H_7O_3^+$ cation population shown in Figure 5a. At $z^+ > 0$, the protonic charge is localized in the limiting solvation structure $H_7O_3^+$ and retains a high intensity at the pivot hydronium. The protonic charge intensities at the two water molecules in $H_7O_3^+$ are higher than those at $z^+ \sim 0$, which corresponds to the increasingly higher Zundel cation population at $z^+ > 0$. In addition, the axial-radial distribution at $z^+ \sim 0$ becomes localized into a few "nodes", which reveals a "frozen" single water wire in which the water positions are tightly constrained in fixed zig-zag positions. At $z^+ > 0$, oxygen positions slowly become less constrained in the axial direction, while at $z^+ > 5$ Å, as the entire proton solvation structure enters the pore, the protonic charge distribution becomes dispersed along the pore. In the $d = 0.95$ nm CNT (7,7) nanopore, the dynamics of the protonic charge distributions is similar to that in CNT (6,6). However, at $z^+ > 0$, a large fraction of the protonic charge is equally localized in two "nodes" with the same axial position, which again indicates a Zundel cation in the orientation perpendicular to the CNT pore.

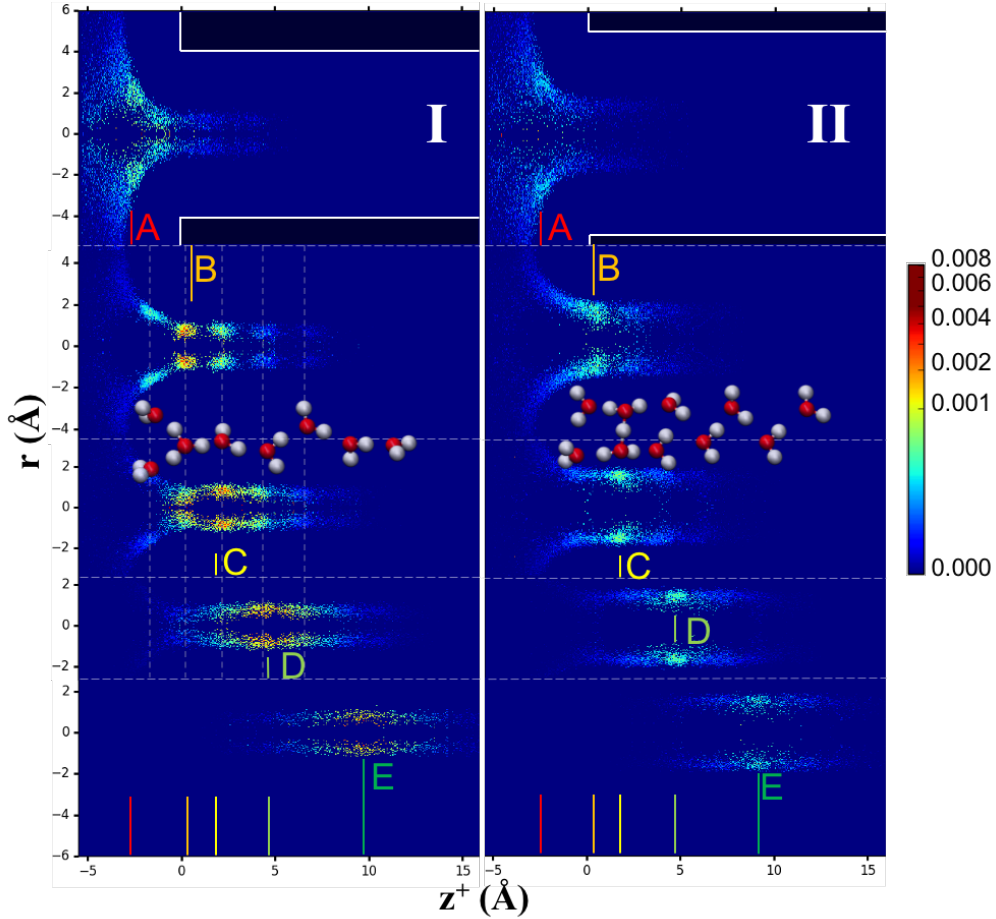

**Figure 9.** The excess protonic charge distribution in I) the $d = 0.81$ nm CNT (6,6) nanopore at $z^+$ = A) -2.5, B) 0.5, C) 2, D) 4.5, E) 9.5 Å and in II) the $d = 0.95$ nm CNT (7,7) nanopore at $z^+$ = A) -2.5, B) 0.5, C) 2, D) 4.5, E) 9.0 Å. The shape and position of CNT nanopores are shown by the black shades in first row of each column. The average excess proton CEC axial position $z^+$ is marked in each figure. Vertical broken lines in I), together with a trajectory snapshot correspondent to I-B), shows the axial position of the "frozen" water wire. A trajectory snapshot correspondent to II-C) shows the Grotthuss shuttling within a Zundel.

3.1.8  Proton shuttle through a frozen water wire via the Grotthuss mechanism

Based on the above analyses, the dynamical behavior of proton transport under confinement could be understood by further analysis of the excess protonic charge distribution. As mentioned above, in the CNT (6,6) nanopore, the protonic charge distribution at $z^+ \sim 0$ (see Figure 9I-B) reveals a "frozen water wire", which includes 1) the core or pivot hydronium, 2) a water molecule towards the bulk, and 3) three water towards the nanopore midpoint. Such a frozen water wire is retained as the proton moves in the $0 < z^+ < 5$ Å region and each member is relatively immobile in its axial and radial positions. As shown in Figure 9I-B-C, the axial positions of each water molecule at $z^+ = 2.0$ Å are the same with those at $z^+ = 0$, and the pivot hydronium position shifts forward by one water molecule at $z^+ \sim 2.0$ Å. The water positions become less confined in the axial direction as the protonic charge propagates into the nanopore. In Figure 9I-D, although the protonic charge distribution is no longer discrete, the frozen water wire positions can still be visualized by the differences in the charge intensities. Such a frozen water wire configuration evidentially shows that the hydrated excess proton shuttles through these water molecules via the Grotthuss mechanism. This is directly correlated to the steep free energy increase as shown in Figure 8. A dispersed or "fuzzy" excess protonic charge distribution, indicating a mobile water wire, corresponds to vehicular diffusion of the hydrated excess proton structure.

In CNT nanopores with larger diameters, as shown in Figure 9II the protonic charge distribution for CNT (7,7), the pivot hydronium and water positions are not consistently sharp and overlaid in Figure 9II-B-D. Thus, the frozen water wire is not present in these larger confined spaces. In the CNT (7,7) nanopore, proton hopping is primarily within the perpendicular oriented Zundel cation, while in $d > 1$ nm CNT nanopores hopping occurs axially within the hydrogen network in the water layer closest to the CNT wall.

Further evidence of a frozen water wire in the narrowest nanotube is observed by examining the evolution of the excess proton solvation structure distribution in Figure 5a. As the proton transports into the CNT (6,6) nanopore, the distorted Zundel and Eigen (or $H_7O_3^+$) cations are alternately the predominant proton solvation structure: the distorted Zundel cation has a larger population at $z^+$ ~1.0 and ~2.5 Å, the $H_7O_3^+$ cation at $z^+ \geq 3.0$ Å, and the Eigen cation at $z^+ \leq 0.5$ Å and $1.5 \leq z^+ \leq 2.0$ Å. The "Eigen–distorted Zundel–Eigen–distorted Zundel–$H_7O_3^+$" sequence, with the aid of the localized protonic charge distribution shown in Figure 9, shows the dynamics of the Grotthuss type proton shuttling through a frozen water wire. In CNT nanopores with larger diameters, there does not exist a similar sequence; instead, the evolution of the proton solvation structure is rather smooth, which corresponds to the findings above of more vehicular diffusion behavior. This combination of multiple types of distributions leads to a more comprehensive understanding of the proton diffusion mechanism in CNTs.

Importantly, the understanding of the proton transport mechanism in different diameters of CNT helps to interpret the diffusion constants reported by Brewer et al[16] (see Figure 9 in Ref. 16). The CNT effective radius $r_{eff}$ determined from the radial distribution in Section 3.1.1 makes the two simulations somewhat equivalent. First, in the $r_{eff}$ = 2.0 Å CNT (6,6) nanopore, the CEC diffusion constant is ~3.9 Å²/ps, and this is due to proton hopping along the nanopore axis exclusively. Second, in the $r_{eff}$ = 2.5 Å nanopore, similar to the $r_{eff}$ = 2.8 Å CNT (7,7) nanopore, Brewer et al also reported a solvated Zundel cation perpendicular to the nanopore axial direction. The CEC diffusion is the lowest in such a nanopore and is ~ 0.1 Å²/ps, since the proton hopping via Grotthuss mechanism is primarily in the radial direction and minimally contributes to proton transport in the CNT axial direction. Third, in the $r_{eff}$ = 3.0, 4.5, 5.0 Å nanopores, i.e., CNT (n,n) (n = 8–10), the excess proton shuttles through the water layer closest to CNT wall via Grotthuss diffusion in both axial and lateral directions, which has a much smaller axial direction proton transport component than that in CNT (6,6). In these nanopores, the hydrogen bond network in the peripheral water layer becomes complex, and the diffusion constant of ~ 0.5 Å$^2$/ps is very similar to that in bulk water.

## 3.2  K$^+$ transport in CNT and compared to hydrated excess proton

### 3.2.1  K$^+$ radial distribution

Potassium cation transport through CNT pores was studied to compare the confinement effect for a different monovalent cation. The effect in CNT with different diameters can be similarly shown first by the K$^+$ radial distribution. As shown in Figure 10, the K$^+$ radial distribution, as well as its evolution along the K$^+$ transport path, differs from that of hydrated excess proton. First, unlike the hydrated excess proton, K$^+$ is consistently distributed at the center of CNT. This is due to the hydrophilic nature of K$^+$ that the ion-dipole interaction between K$^+$ and water is much stronger than the ion-quadrupole interaction between K$^+$ and CNT. Second, as the CNT diameter increases, K$^+$ becomes more confined at the nanopore center. In CNT (8,8), K$^+$ is sandwiched between 5-6 water molecules, and in CNT (9,9), water molecules form a continuous water tube along CNT axial direction and K$^+$ is densely populated at the central cavity. Thus, the K$^+$ position is more confined with a complete first solvation shell.

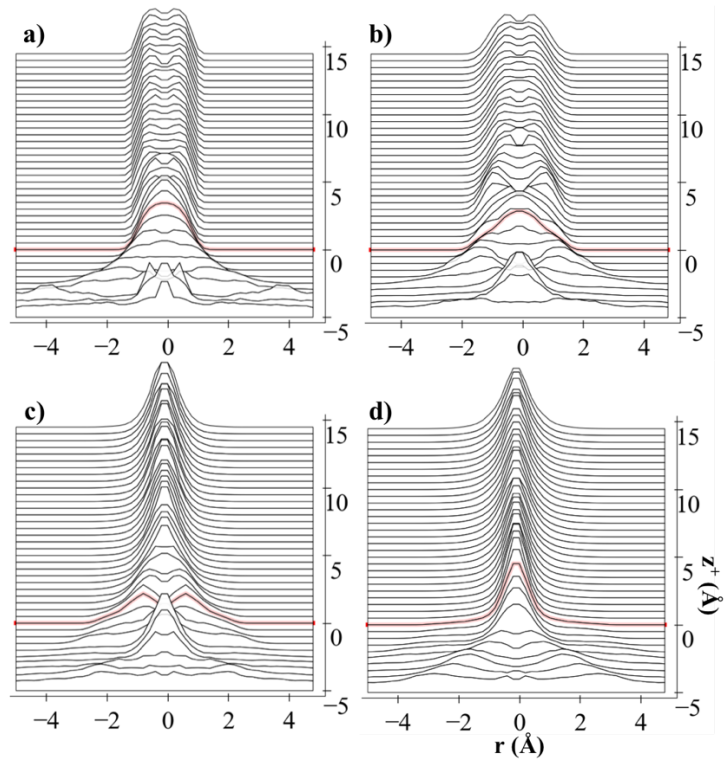

**Figure 10.** Four series of the K$^+$ radial density distributions at -5 < $z^+$ < 15 Å in a) CNT (6,6), b) CNT (7,7), c) CNT (8,8), and d) CNT (9,9) nanopores, respectively.

3.2.2 Free energy of transport of a K$^+$ and "classical" hydronium cation compared with the hydrated excess proton

The free energy profiles of K$^+$ transport through CNTs with different diameters are shown in Figure 11 a)-d) and are compared to those of a hydrated excess proton. Unlike a hydrated excess proton, K$^+$ transport through a CNT nanopore must be solely via vehicular diffusion. As K$^+$ diffuses from bulk water towards a nanopore, it reaches a free energy minimum at $z^+$ = -6.0 Å. Compared to the free energy minimum of proton transport at $z^+ = -3.6$ Å, i.e., the water–graphene interface, the K$^+$ free energy minimum indicates that K$^+$ maintains at least two solvation shells. Next, as K$^+$ moves out of the water bulk, the free energy increases almost linearly without the small flat minimum region as in the proton transfer case. The free energy of removing a K$^+$ from bulk towards the nanopore is higher than that of a proton: at $z^+ = -2.0$ Å, the free energy difference is 2.6, 1.5, 0.7, and 0.8 kcal/mol in CNT nanopore with $d$ = 0.81, 0.95, 1.09, and 1.22 nm, respectively. Furthermore, as K$^+$ move towards the CNT midpoint, the free energy curve converges and is respectively higher than the proton transport in these nanopores. The K$^+$ transport free energy barrier decreases as $d$ increases and is ~2 and ~3 kcal/mol higher than the proton transport free energy at the midpoint in a $d$ < 1 and > 1 nm nanopore, respectively.

A fictitious "classical" H$_3$O$^+$ with only one diabatic state, i.e., a "non-reactive" H$_3$O$^+$ which cannot Grotthuss shuttle protons, was used to calculate a free energy profile via vehicular diffusion. Such a non-reactive H$_3$O$^+$ also does not retain the full physics of the protonic charge delocalization. Thus, the differences in each comparison of free energy curves can provide some insight into proton transport in confinement. In sub-1 nm nanopores, the H$_3$O$^+$ and hydrated excess proton free energy curves are very similar. The H$_3$O$^+$ free energy barrier is ~2 kcal/mol higher in CNT (6,6), and in CNT (7,7) the two barriers are the same. This indicates that the Grotthuss diffusion, rather than vehicular diffusion, plays a role in proton transport through a single water wire. In the presence of double water wires, the Grotthuss diffusion is mainly "rattling" within the Zundel cation in the lateral direction, and it has much less contribution in the axial direction. In wider nanopores, the shape of the excess proton and H$_3$O$^+$ free energy curves are rather different. In CNT (8,8), the free energy increases as H$_3$O$^+$ moves towards CNT midpoint and is ~3 kcal/mol higher than the hydrated excess proton. In CNT (9,9), the $H_3O^+$ free energy at nanopore midpoint is ~2 kcal/mol higher and is almost the same as the free energy minimum outside the nanopore. This implies that the charge delocalization, as well as the Grotthuss diffusion are important in the complex hydrogen network under confinement, even for the larger diameter nanotubes.

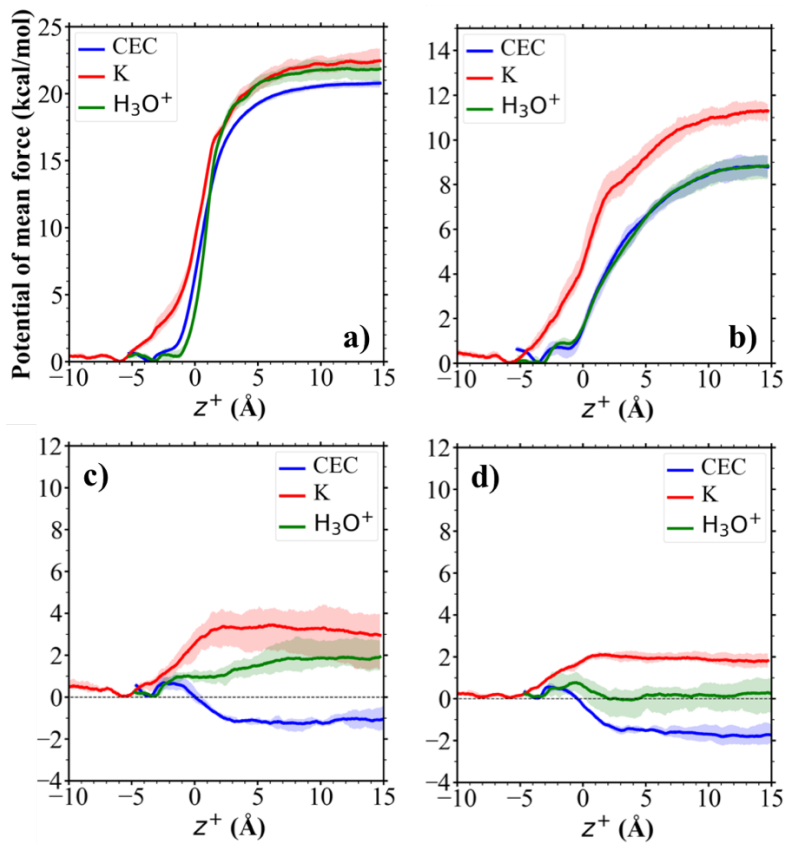

**Figure 11.** The potential energy curves of hydrated excess proton (blue lines), $K^+$ (red lines) and a fictitious non-reactive $H_3O^+$ (green lines) transport in the a) CNT (6,6), b) CNT (7,7), c) CNT (8,8), and d) CNT (9,9) nanopores, respectively.

### 3.2.3  K⁺ induced water density increase

Similar to the proton-induced water density increase effect, the presence of a K⁺ in the nanopore also increases the water density inside CNT (7,7). Before K⁺ transport into the nanopore, a single water wire is established with ~15 water molecules. As K⁺ moves into the nanopore, it creates a short double water wire with the incomplete solvation shell of four water molecules. And the number of water molecules inside the CNT (7,7) nanopore is on average ~18. Similarly, the presence of a non-reactive classical $H_3O^+$ also creates a double water wire with on average ~18 water molecules besides $H_3O^+$. This shows that an excess monovalent positive charge increases the water density by ~ 20 %. Thus, given the ~ 40 % water density increase induced by a hydrated excess proton, nearly 50 % is contributed by the hydrated excess positive charge and the rest by the structural protonic charge delocalization.

### 3.3  Nanopore length, curvature, and chirality

In addition to a variation in diameter, a nanopore could also have different lengths, curvatures, and chirality. The confinement effect with respect to these factors were studied with three CNT nanopore models as shown in Figure 12a: 1) a cylindrical nanopore of $d = 0.81$ nm CNT (6,6) with a length of 5.03 nm, 2) a curved nanopore of CNT with a tangential length of 3.68 nm, and 3) a cylindrical nanopore of $d = 0.82$ nm CNT (1,10) with a length of 2.95 nm. A single water wire, similar to that in CNT (6,6), is formed along each CNT axis.

The free energy profiles of a hydrated excess proton transported through the three CNT nanopores (see Figure 12b) are compared to that using an $l = 2.95$ nm CNT (6,6) nanopore shown in section 3.1.6. The shape of the four free energy curves are almost the same, and the free energy barriers are 20.8, 21.0, 20.9, and 20.3 kcal/mol for CNT (6,6) ($l \sim 3$ nm), CNT (6,6) ($l \sim 5$ nm), curved CNT (6,6) ($l \sim 4$ nm), and CNT (1,10) ($l \sim 3$ nm), respectively. The nearly identical free energy barriers suggest that alternation to the CNT length (longer), curvature, or chirality of a CNT nanopore has little effect on the proton transport free energy barrier. As shown by Levitt, the electrostatic energy between a cation and the lipid membrane increases in a longer nanopore.[77] Thus, according to the X-ray determined DOPC bilayer thickness of 6.31 nm,[78] $F_{electro} = 6.0$ kcal/mol is a better approximation for the cation–lipid electrostatic energy in Section 3.1.6. Wang et al also reported that static water properties in CNT is barely affected by the chirality.[23] Therefore, alternate topologies open new opportunities for synthetic nanoporous membranes and proton channels in transmembrane proteins with equivalent free energy barriers that are determined nearly solely by the diameter rather than length, curvature, or chirality.

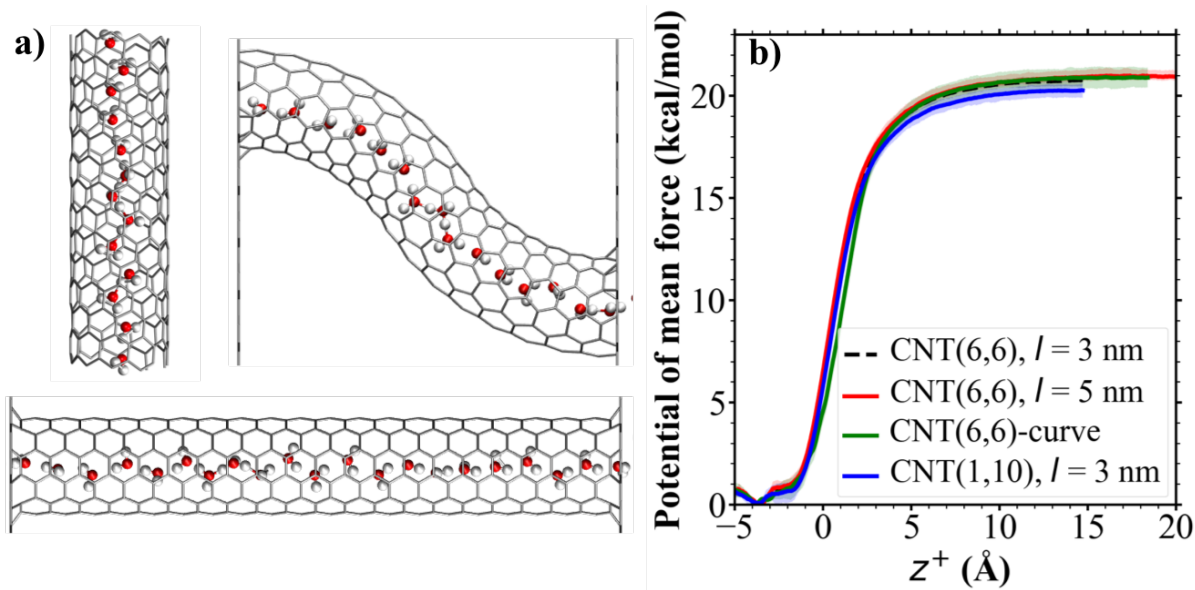

**Figure 12.** a) Side views of hydrated excess proton in single water wires and b) free energy curves of proton transport in a CNT (1,10) ($d$ = 0.82 nm, $l$ ~3 nm), a curved CNT (6,6) ($d$ = 0.81 nm, $l$ ~4 nm) and an elongated CNT (6,6) ($d$ = 0.81 nm, $l$ ~5 nm).

3.4    Nanopores with more complex topology

3.4.1    "Dumbbell" shaped nanopore

Rarely will a hydrophobic nanopore, such as a transmembrane ion channel, have a perfect and simple cylindrical topology. Complex topology with continuously changing diameters may lead to interesting behavior as the result of combinations of various confinement effects. The artificial "dumbbell" shaped CNT nanopores, as shown in Figure 13a, consist of two $d$ = 1.09 nm CNT (8,8) segments and a $d$ = 0.81 nm CNT (6,6) "bottleneck" segment in the middle. The "dumbbell" CNTs were used as a model to study hydrated excess proton transport in a nanopore channel with complex topology. The bottleneck segment, which functions as a gate in the nanopore, creates a free energy barrier of proton transport through the nanopore. Unlike a gate in a protein ion channel, which can open and close the transmembrane pore by side chains motions, the CNT bottleneck segment is rigid, and the barrier solely arises from the confinement effect. The two $d =$ 1.09 nm segments are expected to have minimum influence on the proton transport free energy barrier since the free energy fluctuation of proton transport in an isolated $d = 1.09$ nm CNT (8,8) nanopore is approximately within 1 kcal/mol (Figure 8).

The length of the "dumbbell" shaped nanopores was kept the same as the cylindrical nanopores with different diameters, i.e., $l = 2.95$ nm, which allows the results to be directly compared to those shown earlier. The excess proton CEC radial distribution is plotted versus a series of CEC axial positions to show the features of the excess proton positions under confinement (see Figure 13c). Before and after the proton enters the bottleneck segment, the CEC distributions agree with those in a $d = 1.09$ nm CNT (8,8) pore (Figure 2k) and in a $d = 0.81$ nm CNT (6,6) pore (Figure 2i), respectively. A single water wire inside the bottleneck segment connects water in the two CNT (8,8) segments and allows the proton to shuttle through. The length of the bottleneck segments $l_b$ has a significant influence on the proton transport free energy barrier: for $l_b = 0.5, 1.0$, and 1.5 nm, the free energy barrier is 13.7, 19.5, and 20.5 kcal/mol, respectively. The $l_b = 1.5$ nm "dumbbell" nanopore and the cylindrical CNT (6,6) nanopore have the same proton transport free energy barrier, although the $d = 0.81$ nm segment is shorter by ~1.5 nm. The presence of a 0.5 nm bottleneck, which allows a single water wire of three water molecules, increases the midpoint free energy in a CNT (8,8) nanopore from -1 to ~14 kcal/mol.

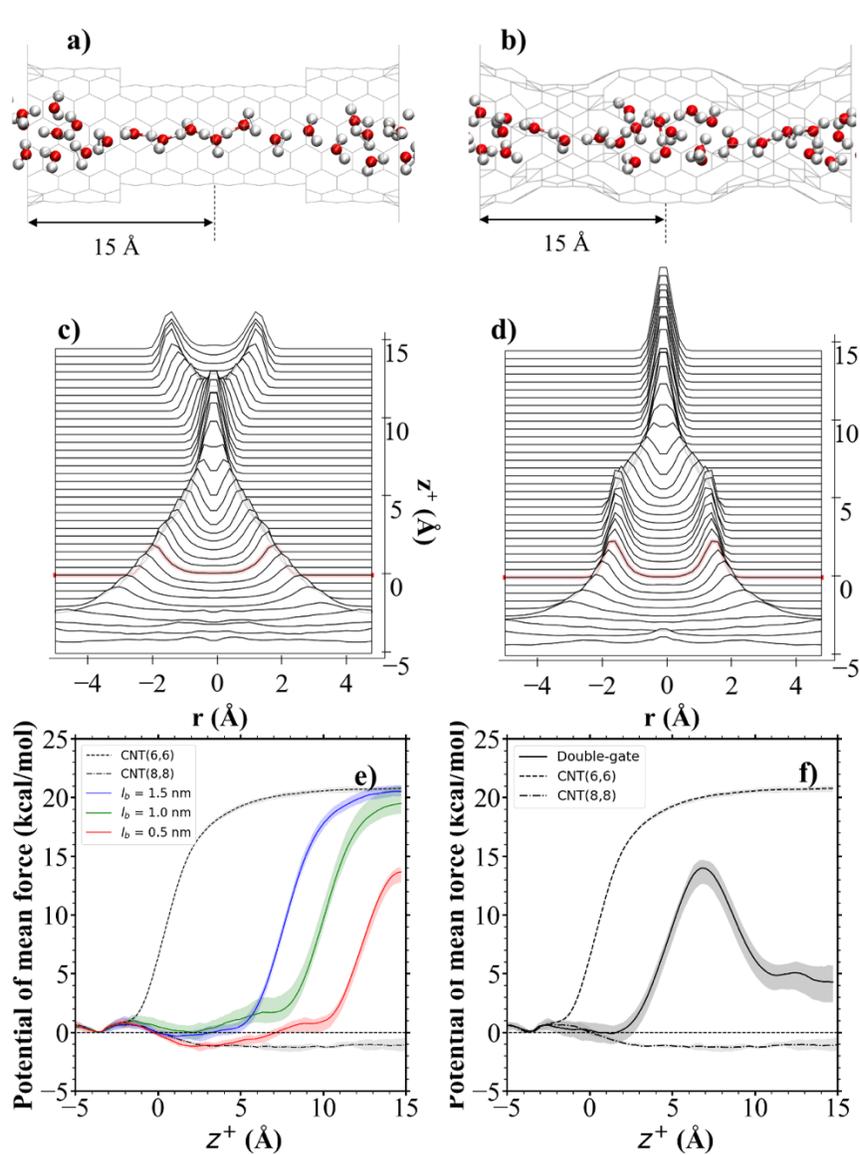

**Figure 13.** a)–b) The sideview and trajectory snapshot of hydrated excess proton and water molecules, c)–d) the excess proton CEC radial density distributions at -5 < $z^+$ < 15 Å, and e)–f) the proton transport free energy curves in the $l_b$ = 1.5 nm "dumbbell" CNT nanopore and the $l_b$ = 0.5 nm "double-gate" nanopore, respectively. The free energy curves of proton transport in "dumbbell" CNT nanopores with $l_b$ = 0.5 (red), 1.0 (green), and 1.5 nm (blue) are also compared with those in the cylindrical $d$ = 0.81 and 1.06 nm CNT nanopores.

### 3.4.2 "Double-gate" shaped nanopore

Another CNT nanopore model with complex topology, named here the "double-gate" shaped nanopore, mimics aspects of an ion channel in a transmembrane protein, e.g., SERCA, which has more than one gate. Due to the continuous transition in the pore topology, the CEC radial distribution changes smoothly and becomes a single-peak distribution in the bottleneck vicinity. Compared to the "dumbbell" shaped CNT nanopore, the double-gate nanopore also consists of an intermediate channel with $d = 12.21$ nm and $l \sim 1$ nm. The excess proton CEC radial distribution at the nanopore midpoint is consistent with that in a CNT (8,8) nanopore. The two $l_b = 0.5$ nm bottlenecks function similarly as the 0.5 nm bottleneck in the "dumbbell" nanopore by creating two 14.0 kcal/mol proton transport free energy barriers. The free energy profile of hydrated excess proton transport through the "double-gate" nanopore, shown in Figure 13f, is the same as that in the $l_b = 1.5$ and 0.5 nm "dumbbell" nanopore, in the region $-5.0 < z^+ < 2.0$ Å and in the bottleneck vicinity, respectively. However, as the proton transports into the intermediate channel, the $F(z^+)$ decreases to $\sim 4.2$ kcal/mol instead of going to 0. This indicates that the confinement effect in the intermediate region, which is bounded by two bottlenecks segments, is greater than that in the entrance channel segment.

### 3.4.3 Frozen water wire at nanopore bottleneck

The distribution of the hydrated excess protonic charge in the "dumbbell" and "double-gate" nanopores reveals the dynamical behavior of the hydrated excess proton solvation structure in the bottleneck vicinity (Figure 14). As the proton approaches the nanopore bottleneck, the protonic charge distributions become spatially discrete (see Figure 14I-D, II-B, II-C, II-D), which is similar to that shown in Figure 14I-B. Furthermore, as the excess proton transports through the bottleneck, the water molecules inside the bottleneck and their neighbors become immobile regardless of the excess proton CEC positions. This indicates a frozen water wire configuration at each of bottlenecks. The $l_b$ = 0.5, 1.0, and 1.5 nm bottlenecks contain a single water wire with approximately 3, 5, and 7 water molecules, respectively. In the longest bottleneck, the protonic charge distribution becomes dispersed as the excess proton moves towards the bottleneck midpoint (see Figure 14I-E), which corresponds to a longer free energy plateau. In the $l$ = 0.5 nm bottleneck, as shown in Figure 14II-B, II-C, II-D, the frozen water wire configuration is retained as the excess proton moves out of the bottleneck. In both "dumbbell" and "double-gate" nanopores, a hydrated excess proton permeates through a nanopore bottleneck by Grotthuss shuttling via the frozen water wires. The vehicular diffusion only occurs either outside a bottleneck or in the bottleneck midpoint, which corresponds to the flat regions in the free energy profile.

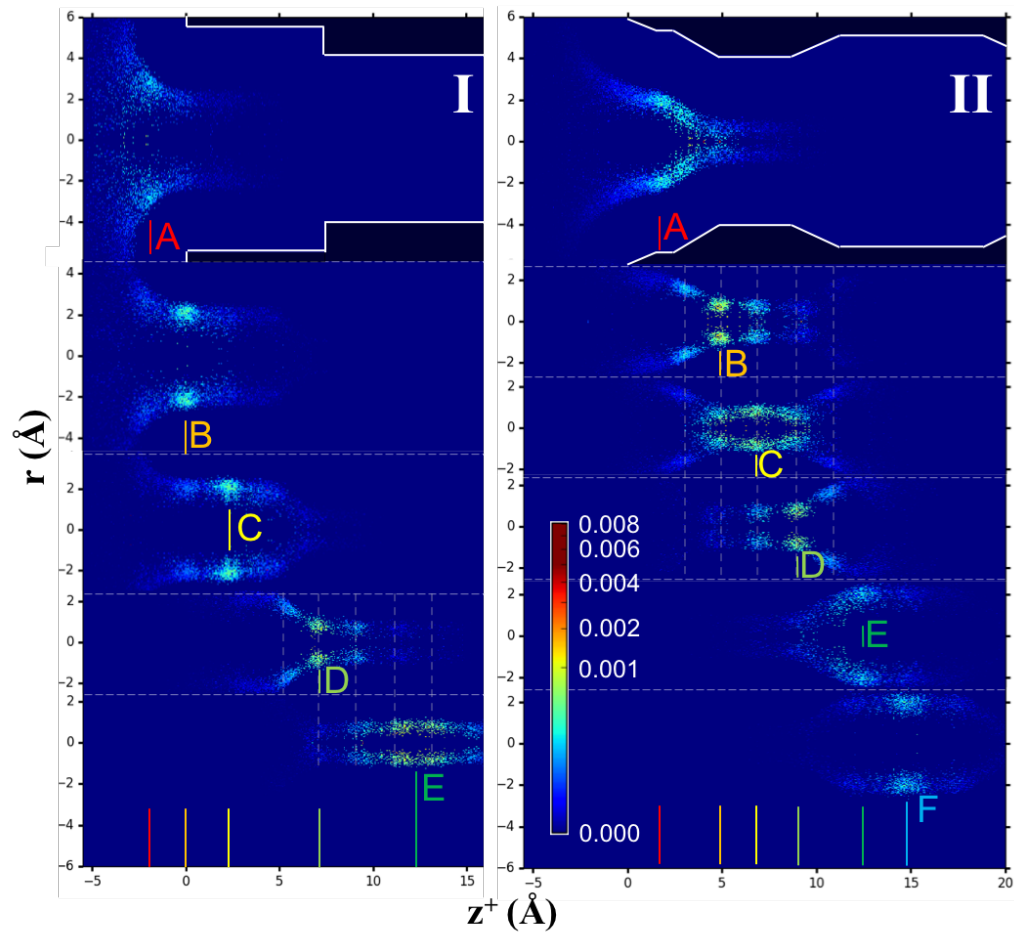

**Figure 14.** The excess charge distribution of hydrated excess proton in I) the "dumbbell" and II) the "double-gate" CNT. The bottleneck lengths are of 1.5 and 0.5 nm, respectively. The shape of CNT is shown in the first row of each series of plots, and several scenarios are plotted to show the dynamics of excess charge distribution. The average CEC axial positions are labeled in each sub-plot and at the bottom and are: $z^+$ = A) -2.0, B) 0, C) 2.5, D) 7.0, and E) 12.0 Å on the left plots and $z^+$ = A) 2.0, B) 5.0, C) 7.0, D) 9.0, E) 13.0 Å and F) 15.0 Å on the right plots, respectively.

## 4. Concluding Remarks

Confinement effects play an important role in hydrated excess proton transport through hydrophobic nanoscale channels. Systematic studies in this work on size-tunable CNT nanopores having simple cylindrical topology have brought new insight into the correlation between the nanopore topology, i.e., diameter, length, curvature and chirality, and the confinement effects. The hydration structure of the hydrated excess proton – the excess proton CEC radial distribution– and the proton transport free energy barrier are primarily determined by the diameter of the hydrophobic nanopore. $K^+$ transport, compared to proton transport, has a higher free energy in each sub-2 nm pore, respectively, which underlines the potential of applying sub-2 nm hydrophobic nanopore in water desalination.

Interestingly, the nanopore confinement effect is largely unperturbed by variations of other CNT topology factors. However, CNT nanopores with multiple different diameters connected in series introduce additional features of the confinement effects. The free energy profiles of proton transport through "dumbbell" nanopores and a longer nanopore shows that the length of a sub-1 nm nanopore, up to a certain length but not beyond that length, determines the proton translocation free energy barrier height. On the other hand, alternations to the nanopore curvature and chirality, without changing the nanopore diameter, have a minor effect on proton the transport free energy profile. By contrast, the intermediate channel bounded between two bottlenecks shows different confinement effects, where the free energy of the hydrated excess proton in such a region is much higher than in a cylindrical pore with the same diameter.

The present work also demonstrates that the MS-EVB model is a powerful tool to resolve in detail the dynamical behavior of the proton solvation structure under confinement through a mapping of the excess protonic charge distribution. In particular, the protonic charge is more localized as the hydrated excess proton enters a nanopore or passes through a $d = 0.81$ nm bottleneck, and the constrained water solvent molecule positions reveal a "frozen" water wire.

Moreover, while proton shuttles through this frozen water wire, Grotthuss diffusion, rather than vehicular diffusion, plays a dominant role in the process. Compared to the free energy profile of a fictitious non-Grotthuss shuttling "classical" $H_3O^+$, the excess protonic charge distribution reveals that Grotthuss diffusion in the axial direction is important in a $d = 0.81$ nm CNT (6,6) nanopore, and much less important in a $d = 0.95$ nm CNT (7,7) nanopore, in which the Grotthuss diffusion is primarily just "rattling" within the interesting orthogonally aligned Zundel cation. These studies thus provide insights into the differences between the ultrafast and ultraslow proton diffusion rates in CNT (6,6) and CNT (7,7). The proton diffusion then becomes more bulk-like in larger-diameter nanopores, but mostly in the water layer adjacent to the hydrophobic CNT wall.

The simulations presented in this work provide multiple predictions and insights into the novel behavior of hydrated excess protons under nanoscale confinement in hydrophobic CNTs, which can be the focus of future detailed experimental studies.

AUTHOR INFORMATION

**Corresponding Author**

* Email: gavoth@uchicago.edu

ACKNOWLEDGMENTS

This work was supported as part of the Center for Advanced Materials for Energy Water Systems (AMEWS), an Energy Frontier Research Center funded by the U.S. Department of Energy (DOE), Office of Science, Basic Energy Sciences (BES). The authors gratefully acknowledge the computational resources provided on the Midway cluster operated by the University of Chicago Research Computing Center. We thank Dr. Yuxing Peng for discussions on the simulations.

**TOC Graphic**

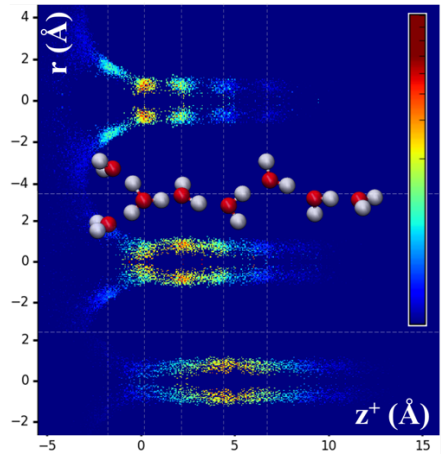

# Water Assisted Proton Transport in Confined Nanochannels

**Supporting Information**

*Xinyou Ma[1], Chenghan Li[1], Alex B. F. Martinson[2] and Gregory A. Voth[1]\**

1. Department of Chemistry, Chicago Center for Theoretical Chemistry, The James Franck Institute, and Institute for Biophysical Dynamics, The University of Chicago, Chicago, IL 60637, United States

2. Materials Science Division, Argonne National Laboratory, Argonne, Illinois 60439, United States

**Animation**

Three animation files, "CNT66-charge_distribution.mov", "CNT77-charge_distribution.mov" and "CNT_2Gates-charge_distribution.mov", show the evolution of excess protonic charge distribution as a hydrated excess proton transporting into and through a CNT (6,6) ($d$ = 0.81 nm), a CNT (7,7) ($d$ = 0.95 nm), and a "double-gate" CNT nanopores, respectively. These animations, plotted with the hydrated excess proton center of excess charge (CEC) distributions in the axial ($z^+$) and radial ($r^+$) dimensions, use the same scale and color scales with those in Figure 9-I, Figure 9-II, and Figure 14-II, respectively. It is worthy to notice that the animation framerate is arbitrarily set to one frame per $z^+$ with a uniform interval of 1 frame/0.5 Å, rather than a realistic (or quasi-realistic) framerate that reflects the actual hydrated excess proton diffusion rate.

URLs:

| | |
|---|---|
| CNT66-charge_distribution.mov: | https://www.youtube.com/watch?v=SvY2-WIvcqA |
| CNT77-charge_distribution.mov: | https://www.youtube.com/watch?v=J-kdbYt7PFY |
| CNT_2Gates-charge_distribution.mov: | https://www.youtube.com/watch?v=eS8403tatPU |